# NOISE AND BELL'S INEQUALITY

DAVID K. FERRY

*School of Electrical, Computer, and Energy Engineering and Center for Solid State Electronic Research, Arizona State University, Tempe, AZ 85287-5706, USA*

LASZLO B. KISH

*Department of Electrical and Computer Engineering, Texas A&M University
College Station, TX 77843-3128, USA*

From the beginning of quantum mechanics, there has been a discussion about the concept of reality, as exemplified by the EPR paradox. To many, the idea of the paradox and the possibility of local hidden variables were dismissed by the Bell inequality. Yet, there remains considerable evidence that this inequality can be violated even by classical systems, so that experiments showing quantum behavior and the violation of the inequality must be questioned. Here, we demonstrate that classical optical polarization experiments based upon noise in the system can be shown to violate the Bell inequality.

## 1. Introduction

Einstein, Podolsky, and Rosen (EPR) [1] set off a discussion over the interpretation of reality within quantum mechanics. One of the important points raised by the EPR experiment is the existence of an objective reality prior to measurement, a point which induced significant dispute within the founding fathers of quantum mechanics. According to the Copenhagen school, reality only appears after measurements of the quantum system, and thus the incompleteness of quantum theory proposed by EPR, is incorrect [2]. The EPR argument was extended via the assertion that, if the quantum description is incomplete, then there must be some hidden variables by which the system can be more fully characterized. It was to deal with this point that Bell provided an inequality satisfied if (local) hidden variables existed [3], yet he showed that quantum systems violated this inequality, specifically because the quantum system in question, entangled states, would be strongly correlated. Bell's results seemed to reinforce the view that quantum mechanics was complete, and that local hidden variables could not be supported (or were not compatible with this completeness). This, of course, depends upon the correctness of the Bell inequality and of the suitability of its application to the question of either the EPR experiment or of the correctness of quantum mechanics. This point has been challenged repeatedly, most recently by Hess and Phillip [4], but Bell's inequality seems to be almost universally accepted today.

Other forms of the inequality put forward by Bell have been proposed and possible experimental tests suggested [5,6,7]. But, it remained for Aspect, and his colleagues to actually begin to make measurements on the polarization of entangled pairs of photons [8,9,10]. By measuring the correlation between two polarizers, which could be moved during the flight of the photons, they were able to establish a violation of another variant of the Bell inequality, the CHSH inequality [5]. What is important here is the assumption that the Bell inequalities establish a test that distinguishes between an extra parameters (so-called hidden variables) theory versus quantum mechanics, with the former being constrained by Bell's inequalities and presumably being classical. Hence, the presumed conclusion from the experiment is that the quantum process used to create entangled photons by Aspect *et al.*, and the resulting violations of Bell's inequality that are observed, are clear support for the correctness of quantum mechanics, particularly as espoused in the so-called Copenhagen interpretation. But, can these measurements really distinguish between quantum entanglement and classical correlation?

It is not at all clear that Bell's inequality can distinguish between classical and quantum theories, a point reinforced recently by Hess *et al.* [11]. Brody earlier demonstrated that a pair of phase correlated (classical) pendulums could provide a violation of Bell's inequality [12]. Even earlier, it had been shown that a three dimensional harmonic oscillator violated a discretized version of the Bell inequality [13]. More recently, a particle-based event-by-event simulation has been used to demonstrate exact agreement with quantum theory [14], a result which is not supposed to exist in Bell's worldview. These results would suggest then that experiments such as those of Aspect *et al.* cannot be used to examine the correctness of quantum mechanics. The question really is whether or not strongly correlated classical fields will also violate Bell's inequality. In a recent paper [15], it was shown that this was the case, in keeping with the results mentioned above, and that Aspect's experiments could not distinguish between a strongly correlated classical set of photons and a pair of quantum entangled photons. In this regard, it is important to note that there is nothing in Bell's inequality, nor in the CHSH version, that restricts the applicability to single pairs of photons.

Let us expand upon that point. In the development of these inequalities, assumptions have been made. The most important of these, which Ghirardi [16] refers to as Bell's view of reality, is that one assumes that the probability of two events is the product of the individual probabilities—that is one assumes that the two events are uncorrelated. But, with entangled quantum states, this is not true, and inequalities are violated primarily because the quantum system does not conform to the assumptions. Thus, there is the inferred assumption that quantum states of interest are strongly correlated and classical ones will not. As pointed out though, it is possible to generate correlated classical systems, which will violate the inequality as well because they also don't meet the assumptions in the formulations of the inequalities.

Here, we go further and ask the question as to how much correlation is needed to violate these inequalities. We suggest that a set of polarized waves can be generated in which the polarization is randomly selected by noise. We then show that reasonable measurements of the polarization of these correlated beams can yield measurements which violate the CHSH inequality used by Aspect *et al.* As before, we arrive at the conclusion that measurements such as these cannot distinguish between classical correlation and quantum entanglement. It is important to point out, however, that we do not question the correctness of quantum mechanics, only the ability of some experiments to demonstrate that correctness, particularly when based upon the Bell inequalities.

## 2. The System

Let us consider a particular type of noise, random telegraph waves (RTW). This noise is characterized as a signal which takes two values, either +1 or -1, with equal probability. We are going to assume that we have two independent sources of classical electromagnetic wave for which the polarization can be switched from vertical to horizontal by this RTW. When the RTW signal is +1, the polarization is set to vertical (V). When the RTW is -1, the polarization is set to horizontal (H). It is important to note that the nature of this noise source is not important to the argument. Any noise source with an average value of 0, so that it has equally probably excursions to positive and negative values would work as well, merely by selecting V for positive and H for negative values. However, the use of the RTW simplifies the argument, and this is the particular reason for this choice.

We shall use two such RTWs to create the two polarized beams necessary for the experiment. In particular, we define the conditional probability N(1|1) as the probability that noise source one is +1 given that noise source two is +1. Then, it can be easily shown that the correlation coefficient is given by

$$r = 1 - 2N(1|-1) . \tag{1}$$

Similarly, we define the conditional probability $N(1|-1)$ as the probability that noise source one is +1 given that noise source two is -1. We note that the relations

$$\begin{aligned} N(1|1) &= N(-1|-1) \\ N(-1|1) &= N(1|-1) \end{aligned} \tag{2}$$

hold quite generally. These lead to the important quantities

$$\begin{aligned} N(1|1) &= N(-1|-1) = \frac{1+r}{2}, \\ N(-1|1) &= N(1|-1) = \frac{1-r}{2}, \end{aligned} \tag{3}$$

since

$$N(1|1) + N(1|-1) = 1. \tag{4}$$

As discussed, the two noise sources are used to polarize the two electromagnetic waves, and these are detected by the experiment's polarizers.

## 3. Creating the Measurement

We now want to use two independent observers to monitor the orientations of the polarization of the two classical waves. Now, the electric field of any polarized wave in an arbitrary polarization direction can still be decomposed into two orthogonal components, say $E_V$ and $E_H$, such that the sum of the squares of these two components is equal to the square of the field of the polarized wave. In fact, this is the normal manner in which the polarization is measured; the wave is projected onto detectors on these two orthogonal axes, and the polarization angle inferred from the detected amplitudes with simple geometry [17,18,19]. With the principle axes of the detector used to designate the "vertical" and "horizontal" coordinates, we may write the vector electric field as

$$E = \mathbf{a}_V \cos\theta + \mathbf{a}_H \sin\theta \tag{5}$$

for our normalized wave amplitude. Here, the angle $\theta$ has two components: the relative orientation $\Phi$ of the detector with respect to the laboratory coordinates and the actual angle $\xi$ of the polarization with respect to these latter coordinates.

For the experiment here, we have two waves, so we will need two detectors, which we will label *A* and *B*. From Bell's arguments, we can define the correlation field for coincident measurements as

$$p(A,B) = E_A(\theta_A) \cdot E_B(\theta_B) \ . \tag{6}$$

Such a definition, in consideration of Bell's inequality, has been referred to as a form of local reality, in that it is assumed that each measurement can be treated as an independent quantity without any long-range nonlocal considerations. The probability, as used by Aspect *et al.* [9], is based upon the intensity of the signal, which arises from one-half the square magnitude of the correlation field, as (here, we have no difference in the vertical or horizontal coincidence signals)

$$\begin{aligned} P(V,V) = P(H,H) &= \frac{1}{2} p^2(A,B) \\ &= \frac{1}{2}\left[\cos\theta_A \cos\theta_B + \sin\theta_A \sin\theta_B\right]^2 \ . \\ &= \frac{1}{2}\cos^2(\theta_A - \theta_B) \end{aligned} \tag{7}$$

Now, however, we have to introduce the correlation coefficient for the original noise generated polarizations. The result (4) arises only if the original polarizations are identical, and to account for the noise, we replace (4) with

$$P(V,V) = P(H,H) = \frac{1}{2}\cos^2(\theta_A - \theta_B)\frac{1+r}{2} \ . \tag{8}$$

In a similar manner, if we set the two polarizations opposite, we arrive at

$$P(V,H) = P(H,V) = \frac{1}{2}\sin^2(\theta_A - \theta_B)\frac{1-r}{2} \ . \tag{9}$$

Aspect *et al.* [9] have given the probability of obtaining the correlation result in terms of (5) and (6) with the estimator

$$\begin{aligned} E(A,B) &= P(V,V) + P(H,H) - P(V,H) - P(H,V) \\ &= \frac{1}{2}\cos[2(\theta_A - \theta_B)] + \frac{r}{2} \end{aligned} \ . \tag{10}$$

Now, the most general form of Bell's inequality has been given by Clauser *et al.* [5] as [20], and the one used by Aspect *et al.*,

$$|E(A,B) - E(A,D)| + |E(C,B) + E(C,D)| \leq 2 \ . \tag{11}$$

If we now take the angles as $\theta_A = 0°$, $\theta_B = 22.5°$, $\theta_C = 45°$, $\theta_D = 67.5°$, then we find, using (7) to evaluate the estimators, that

$$\begin{aligned} result &= |E(A,B) - E(A,D)| + |E(C,B) + E(C,D)| \\ &= \frac{1}{2}|\cos(-45) - \cos(135)| + \frac{1}{2}|\cos(45) + \cos(-45) + 2r| \\ &= \frac{1}{2}\left|\frac{1}{\sqrt{2}} + \frac{1}{\sqrt{2}}\right| + \frac{1}{2}\left|\frac{1}{\sqrt{2}} + \frac{1}{\sqrt{2}} + 2r\right| = 1.414 + r. \end{aligned} \quad (12)$$

It is now clear that a violation of (11) will occur whenever $r > 0.656$. Hence, even noise-correlated components of the classical waves will lead to a violation of the Bell inequality (or its equivalents). One should note that there will be a set of angles for which the anti-correlation will also yield a violation of the inequality.

### 4. Discussion

The generally regarded value for the result of (8) for quantum systems is 2.828, which is larger than the possible result of (9). However, the angles used here are the common ones, and no search for a set of angles which will maximize (9) has been undertaken. However, De Raedt *et al.* [14] have shown how even the value in (9) can be exceeded in some cases for particle-like event-based simulations, which reproduce quantum theory. In their view, one must be very careful in defining probability spaces in connection with experiments as the latter may well require the careful definition of a set of generalized coordinates that can be tracked throughout the entire set of experiments.

For sure, the assumption (3), if we are to treat it as a statement of local reality with no fuzzy action at a distance, suggests that very well defined probability spaces for events *A* and *B* must exist, and they must be independent if this is to be the joint probability. Yet, the construction of well-correlated light waves, in terms of their polarizations, seems to contradict this assumption. When we have two measured variables, each with their own probability space, the covariance of the two measurements is given by the joint probability minus the product of the individual expectations; e.g., the joint probability minus (3) [21]. The two measures are independent only if this covariance vanishes.

We can draw a couple of conclusions from the above classical construction and measurement. First, it is probably important to note that expressions such as (3) are simply statements about probability spaces, and should not be construed to have any connection with the concept of reality without more extensive discussions of the experimental conditions as well as the systems which will actually be measured. Hence, in this context it is not suitable to claim that expressions such as these are really statements about local (or nonlocal) reality, since one must be careful in defining to what reality actually refers.

Previously, it was shown that measurements such as those made by Aspect *et al.* [10] really cannot distinguish between entanglement due to the quantum process by which a pair of photons is created and classical correlation imposed upon two photon beams. Here, we have used parametric noise to induce real correlations between the polarizations of two classical source beams, and this has led to a violation of the inequalities. Thus, one might ask if measurements such as those discussed see anything more than correlated noise in the system.

Finally, it must be remarked that it cannot be ruled out that some quantum effects have slipped into the system, especially in the low photon density limit. But, this is unlikely, and it is also unlikely that, should it have occurred, it would have led to entangled photons.